\documentclass[aps,showpacs,manuscript,12pt]{revtex4}
\usepackage{amssymb}
\usepackage{amsmath}
\usepackage{graphicx}

\begin{document}

\title{Kinetic description of classical matter infalling in black holes}
\author{Piero Nicolini$^{1}$ and Massimo Tessarotto$^{1,2}$}
\affiliation{$^{1}$Department of Mathematics and Informatics,
University of Trieste, Trieste, Italy \\
$^{2}$ Consortium for Magnetofluid Dynamics, Trieste, Italy}

\begin{abstract}
A popular aspect of black holes physics is the mathematical
analogy between their laws, coming from general relativity and the
laws of thermodynamics. The analogy is achieved by identifying a
suitable set of observables, precisely: \emph{(a)} $E=M$ (being
$E$ the thermodynamic free energy and $M$ the mass of the BH),
\emph{(b)} $T=\alpha \kappa $ (with $T$ the absolute
temperature, $\kappa $ the so-called surface gravity on event horizon and $%
\alpha $ a suitable dimensional constant) and \emph{(c)}
$S=(1/8\pi \alpha )A $ (where $S$ is the thermodynamic entropy of
the black hole and $A$ the surface of the event horizon). However,
despite numerous investigations and efforts spent on the subject,
the theoretical foundations of such identifications between
physical quantities belonging to apparently unrelated frameworks
are not yet clear. The goal of this work is to provide the
contribution to the black hole entropy, coming from matter in the
black hole exterior. We propose a classical solution for the
kinetic description of matter falling into a black hole, which
permits to evaluate both the kinetic entropy and the entropy
production rate of classical infalling matter at the event
horizon. The formulation is based on a relativistic kinetic
description for classical particles in the presence of an event
horizon. An H-theorem is established which holds for arbitrary
models of black holes and is valid also in the presence of
contracting event horizons.
\end{abstract}

\maketitle


\section{Introduction}

The remarkable mathematical analogy between the laws of
thermodynamics and black hole (BH) physics following from
classical general relativity still escapes a complete and
satisfactory interpretation. In particular it is not yet clear
whether this analogy is merely formal or leads to an actual
identification of physical quantities belonging to apparently
unrelated
framework. The analogous quantities are $E\leftrightarrow M$, $%
T\leftrightarrow\alpha\kappa$ and
$S\leftrightarrow(1/8\pi\alpha)A$, where $A $ and $\kappa$ are the
area and the surface gravity of the BH, while $\alpha$ is a
constant. A immediate hint to believe in the thermodynamical
nature of BH comes from the first analogy which actually regards a
unique physical quantity: the total energy. However, at the
classical level there are obstacles to interpret the surface
gravity as the BH temperature since a perfectly absorbing medium,
discrete or continuum, which is by definition unable to emit
anything, cannot have a temperature different from absolute
zero. A reconciliation was partially achieved by in 1975 by Hawking \cite%
{Hawking 1975}, who showed, in terms of quantum particle pairs
nucleation, the existence of a thermal flux of radiation emitted
from the BH with a
black body spectrum at temperature $T=\hbar\kappa/2\pi k_{B} $ (\textit{%
Hawking BH radiation model}). \noindent The last analogy results
the most intriguing, since the area $A$ should essentially be the
logarithm of the number of microscopic states compatible with the
observed macroscopic state of the BH, if we identify it with the
Boltzmann definition. In such a context, a complication arises
when one strictly refers to the internal microstates of the BH,
since for the infinite red shift they are inaccessible to an
external observer. An additional difficulty with the
identification $S\leftrightarrow(1/8\pi\alpha)A$, however, follows
from the BH radiation model, since it predicts the existence of
contracting BH for which the radius of the BH may actually
decrease. To resolve this difficulty
a modified constitutive equation for the entropy was postulated \cite%
{Bekenstein 1973,Bekenstein 1974}, in order to include the
contribution of the matter in the BH exterior, by setting
\begin{equation}
S^\prime = S+\frac{1}{4}k\frac{c^3 A}{G \hbar },  \label{Beck}
\end{equation}
($S^\prime$ denoting the so-called \textit{Bekenstein entropy})
where $S$ is
the entropy carried by the matter outside the BH and $S_{bh}\equiv\frac{1}{4}%
k\frac{c^3 A}{G \hbar }$ identifies the contribution of the BH. As
a consequence a generalized second law
\begin{equation}
\delta S^\prime\geq 0  \label{second-law}
\end{equation}
was proposed \cite{Bekenstein 1973,Bekenstein 1974} which can be
viewed as nothing more than the ordinary second law of
thermodynamics applied to a system containing a BH. From this
point of view one notices that, by assumption and in contrast to
the first term $S$, $S_{bh}$ cannot be interpreted, in a proper
sense, as a physical entropy of the BH, since, as indicated above,
it may decrease in time.

This approach however is unconvincing since the precise definition
and underlying statistical basis both for $S$ and $S_{bh}$ remain
obscure. Thus a fundamental problem still appears their precise
estimates based on suitable microscopic models. Since the
evaluation of $S_{bh}$ requires the knowledge of the internal
structure of the event horizon (excluding for causality the BH
interior), the issue can be resolved only in the context of a
consistent formulation of quantum theory of gravitation
\cite{Bekenstein
1975,Hawking 1976}. This can be based, for example, on string theory \cite%
{Reviews 1998} and can be conveniently tested in the framework of
semiclassical gravity \cite{Nicolini 2001,Nicolini 2002}.
Regarding, instead the entropy produced by external matter $S$,
its evaluation depends on the nature, not yet known, of the BH.
However, even if one regards the BH as a purely classical object
surrounded by a suitably large number of classical particles its
estimate should be achievable in the context of classical
statistical mechanics.

In statistical mechanics the ``disorder'' characterizing a
physical system, classical or quantal, endowed by a large number
of permissible microstates, is sometimes conventionally measured
in terms of the so-called Boltzmann entropy $S_B=K\ln W.$ Here $K$
is the Boltzmann constant while $W$ is a suitable real number to
be identified with the total number of microscopic complexions
compatible with the macroscopic state of the system, a number
which generally depends on the specific micromodel of the system.
Therefore, paradoxically, the concept of Boltzmann entropy does
not rely on a true statistical description of physical systems,
but only on the classification
of the internal microstates (quantal or classical). As is well known\cite%
{Chakrabarti 1976}, $S_B$ can be axiomatically defined, demanding
(i) that it results a monotonic increasing function of $W$ and
(ii) that it satisfies the entropy additivity law $S_B(W_1
W_2)=S_B(W_1)+S_B(W_2)$. Boltzmann entropy plays a crucial role in
thermodynamics where (i) and (ii) have their corresponding laws in
the entropy nondecreasing monotonicity and additivity. Since in
statistical mechanics of finite system it is impossible to satisfy
both laws exactly, the definition of $S_B$ is actually conditioned
by the requirement of considering systems with $W>>1$ (large
physical systems).

An alternate definition of entropy in statistical mechanics is the
one given by the Gibbs entropy, in turn related to the concept of
Shannon information entropy. In contrast to the Boltzmann entropy,
this is based on a statistical description of physical systems and
is defined in terms of the probability distribution of the
observable microstates of the system. In many cases it is
sufficient for this purpose to formulate a kinetic description,
and a corresponding kinetic entropy, both based on the
one-particle kinetic distribution function. In particular, this is
the case of classical many-particle systems, consisting of weakly
interacting ultra relativistic point particles, such as those
which may characterize the distribution of matter in the immediate
vicinity of the BH exterior.

These issue have motivated a recent research effort on the subject
by the present authors \cite{Nicolini2006}. The primary goal has
been, in particular, to provide an explicit expression for the
contribution $S$, which characterizes Bekenstein law (\ref{Beck}),
to be evaluated in terms of a suitable kinetic entropy, and to
estimate the corresponding entropy production rate due to
infalling matter at the BH event horizon. In addition we intend to
establish an H-theorem for the kinetic entropy which holds, in
principle, for a classical BH characterized by event horizons of
arbitrary shape and size and even in the presence of BH implosions
or slow contractions. This is obtained in the framework where the
classical description of outside matter and space is a good
approximation to the underlying physics.

\section{Kinetic description of infalling matter}

The basic assumption is that the matter falling into the BH is
formed by a system $S_{N}$ of $N$ classical point particles moving
in a classical spacetime and described by Hamiltonian dynamics,
while the event horizon can be treated as a classical absorbing
porous wall. We adopt for this purpose a covariant kinetic
formalism taking into account the presence of an event horizon.
The evolution of such a system is well known and results uniquely
determined by the classical equations of motion, defined with
respect to an arbitrary observer $O$. To this purpose let us
choose $O$, without loss of generality, in a region where space
time is (asymptotically) flat, endowing
with the proper time $\tau $, with $\tau $ assumed to span the set $%
I\subseteq R$ (observer's time axis). Each particle is described
by the canonical state $\mathbf{x}$, which in the case of point
particles spans the $8-$dimensional phase space $\Gamma ,$ where
${\mathbf{x}}=\left( r^{\mu },p_{\mu }\right) $. Moreover, its
evolution is prescribed in terms of a suitable relativistic
Hamiltonian $H=H\left( \mathbf{x}\right) $ so that the canonical
state ${\mathbf{x}}=\left( r^{\mu },p_{\mu }\right) $ results
parameterized in terms of the world line arc length $s$ (see
\cite{Synge 1960}). As a consequence, requiring that $s=s(\tau )$
results a strictly monotonic function it follows that, the
particle state can be also parameterized in terms of the
observer's time $\tau .$ To obtain the a kinetic description for
such a system we require, as usual, $N\gg 1.$ In addition, it is
assumed that interactions between point particles of $S_{N}$ take
place only via a mean-field Hamiltonian $H\left( \mathbf{x}\right)
$ and hence that $S_{N}$ can be identified with a weakly
interacting relativistic gas. For $S_{N}$ we introduce the kinetic
distribution function for the observer $O$, $\rho
_{G}(\mathbf{x})$, defined as follows
\begin{equation}
\rho _{G}({\mathbf{x}})\equiv \rho ({\mathbf{x}})\delta (s-s(\tau ))\delta (%
\sqrt{u_{\mu }u^{\mu }}-1)
\end{equation}%
where $\rho \left( \mathbf{x}\right) $ is the conventional kinetic
distribution function in the $8-$dimensional phase space, to be
assumed
suitably smooth and summable in $\Gamma $ (\emph{Assumption }$\alpha $\emph{%
\ of regularity})$.$ Notice that the Dirac deltas here introduced
must be intended as \emph{physical realizability equations}. In
particular the condition placed on the arc length $s$ implies that
the particle of the system is parameterized with respect to
$s({\tau })$, i.e., it results functionally dependent on the
proper time of the observer; instead the constraints placed on
$4$-velocity implies that $u^{\mu }$ is a tangent vector to a
timelike geodesic. The \emph{event horizon} of a classical BH is
defined by the surface $r_{H}$ specified by the equation
\begin{equation}
R(x)=r_{H}
\end{equation}%
where $x$ denotes a point of the space time manifold, while $R(x)$
reduces to the radial coordinate in the spherically symmetric
case.

According to a classical point of view, let us now assume that the
particles are "captured" by the BH (i.e., for example, they
effectively disappear for the observer since their signals are red
shifted in such a way that they cannot be anymore detected
{\cite{Wald 1984}}) when they reach a suitable surface $\gamma $
(\emph{capture surface}) of equation
\begin{equation}
R_{\epsilon }(x)=r_{\epsilon }.
\end{equation}%
Here $r_{\epsilon }=(1+\epsilon )r_{H}$, while $\epsilon >0$
depends on the detector and the $4-$momentum of the particle. The
presence of the BH event horizon is taken into account by defining
suitable boundary conditions for the kinetic distribution function
on the hypersurface $\gamma ,$ to be treated as an effective
absorbing porous wall. For this purpose, we distinguish between
incoming and outgoing distributions on $\gamma $, $\rho
_{G}^{-}(\mathbf{x})$ and $\rho _{G}^{+}(\mathbf{x})$
corresponding
respectively to $n_{\alpha }u^{\alpha }\leq 0$ and $n_{\alpha }u^{\alpha }>0$%
, where $n_{\alpha }$ is a locally radial outward $4-$vector.
Therefore, the boundary conditions on $\gamma $ are specified as
follows
\begin{eqnarray}
\rho _{G}^{-}({\mathbf{x})} &\equiv &\rho ({\mathbf{x}})\delta
(s-s(\tau
))\delta (\sqrt{u_{\mu }u^{\mu }}-1)  \label{bc-1} \\
\rho _{G}^{+}({\mathbf{x}}) &\equiv &0.  \label{bc-2}
\end{eqnarray}%
It follows that it is possible to represent the kinetic
distribution function in the whole space time manifold in the form
\begin{equation}
\rho _{G}\left( \mathbf{x}\right) =\rho _{G}^{-}\left(
\mathbf{x}\right) +\rho _{G}^{+}\left( \mathbf{x}\right)
\end{equation}%
where
\begin{eqnarray}
\rho _{G}^{\pm }({\mathbf{x}}) &\equiv &\rho ({\mathbf{x}})\delta
(s-s(\tau
))\delta (\sqrt{u_{\mu }u^{\mu }}-1)\times   \nonumber \\
&&\times \Theta ^{\pm }(R_{\epsilon }(x)-r_{\epsilon }(s(\tau )))
\end{eqnarray}%
with $\Theta ^{\pm }$ respectively denoting the strong and the
weak Heaviside functions
\begin{equation}
\Theta ^{-}(a)=\left\{
\begin{array}{ccc}
1 & for & a\geq 0 \\
0 & for & a<0.%
\end{array}%
\right.
\end{equation}%
and
\begin{equation}
\Theta ^{+}(a)=\left\{
\begin{array}{ccc}
1 & for & a>0 \\
0 & for & a\leq 0.%
\end{array}%
\right.
\end{equation}%
In the sequel we shall introduce the hypothesis that the
distribution $\rho _{G}^{-}({\mathbf{x})}$ has a compact support
(\emph{Assumption }$\beta $).

It is important to stress that in the definition of the boundary
conditions no detailed physical model is actually introduced for
the particle loss mechanism, since all particles are assumed to be
captured on the same hypersurface $\gamma $, independent of their
mass, charge and state. This provides a classical loss model for
the BH event horizon.

Let us now consider the evolution of the kinetic distribution
function $\rho _{G}(\mathbf{x})$ in external domains, i.e. outside
the event horizon. Assuming that binary collisions are negligible,
or can be described by means of a mean field, and provided that
the phase space volume element is conserved, it follows the
collisionless Boltzmann equation, or the Vlasov equation in the
case of charged particles \cite{Tessarotto2004},
\begin{equation}
\frac{ds}{d\tau }\left\{ \frac{dr^{\mu }}{ds}\frac{\partial \hat{\rho}(%
\mathbf{x})}{\partial r^{\mu }}+\frac{dp_{\mu }}{ds}\frac{\partial \hat{\rho}%
(\mathbf{x})}{\partial p_{\mu }}\right\} =0  \label{Liouville
equation}
\end{equation}%
with summation understood over repeated indexes, while $\widehat{\rho }(%
\mathbf{x})$ denotes $\rho _{G}(\mathbf{x})$ evaluated at $%
r^{0}=r^{0}(s(\tau ))$ and $p_{0}=m\left\vert \frac{\partial r_{0}(s)}{%
\partial s}\right\vert _{s=s(\tau )}.$ This equation resumes the
conservation of the probability in the relativistic phase space in
the domain external to the event horizon. Invoking the Hamiltonian
dynamics for the system of particles, the kinetic equation takes
the conservative form
\begin{equation}
\frac{ds}{d\tau }\left[ \hat{\rho}({\mathbf{x}}),H\right]
_{\mathbf{x}}=0.
\end{equation}

\section{BH statistical entropy and H-theorem}

Let us now introduce the appropriate definition of kinetic entropy
$S(\rho )$ in the context of relativistic kinetic theory. We
intend to prove that in the presence of the BH event horizon it
satisfies an $H$ theorem.

The concept of entropy in relativistic kinetic theory can be
formulated by direct extension of customary definition given in
nonrelativistic setting \cite{Israel 1963,Cercignani 1975,De Groot
1980}. For this purpose we introduce the notion of kinetic
entropy, measured with respect to an observer endowed with proper
time $\tau ,$ as follows
\begin{equation}
S(\rho )=-P{\int\limits_{\Gamma }}d{\mathbf{x}}(s)\delta (s-s(\tau ))\delta (%
\sqrt{u_{\mu }u^{\mu }}-1)\rho ({\mathbf{x}})\ln \rho
({\mathbf{x}}), \label{Kentropy}
\end{equation}%
where $\rho (\mathbf{x})$ is strictly positive and, in the
8-dimensional
integral, the state vector $\mathbf{x}$ is parameterized with respect to $s$%
, with $s$ denoting an arbitrary arc length. Here $P$ is the
principal value of the integral introduce in order to exclude from
the integration domain the subset in which the distribution
function vanishes. Hence $S(\rho )$ can also be written:
\begin{equation}
S(\rho )=-P{\int\limits_{\Gamma }}d{\mathbf{x}}(s)\delta (s-s(\tau
))\rho _{1}({\mathbf{x}})\ln \rho (\mathbf{x}),  \label{kinetic
entropy}
\end{equation}%
where $\rho _{1}(\mathbf{x})$ now reads%
\begin{equation}
\rho _{1}({\mathbf{x}})=\Theta (r(s)-r_{\epsilon }(s))\delta
(\sqrt{u_{\mu }u^{\mu }}-1)\rho ({\mathbf{x}}(s)).
\end{equation}%
In the sequel $S(\rho )$ will be denoted as \emph{BH classical
entropy}. Differentiating with respect to $\tau $ and introducing
the invariant volume element $d^{3}\mathbf{r}d^{3}\mathbf{p}$, the
entropy production rate results manifestly proportional to the
area $A$ of the event horizon and reads
\begin{equation}
\frac{dS(\rho )}{d\tau }\equiv \frac{dS_{1}}{d\tau }+\frac{dS_{2}}{d\tau }=-P%
{\int\limits_{\Gamma ^{-}}}d^{3}{\mathbf{r}}d^{3}{\mathbf{p}}F_{{r}{%
r_{\epsilon }}}\left[ \delta \left( r-r_{\epsilon }\right)
\hat{\rho}\ln \hat{\rho}\right] ,  \label{entro}
\end{equation}%
where $F_{{r}{r_{\epsilon }}}$ is the characteristic integrating
factor
\begin{equation}
F_{{r}{r_{\epsilon }}}\equiv \frac{ds(\tau )}{d\tau }\left( \frac{dr}{ds}-%
\frac{dr_{\epsilon }}{ds}\right) .
\end{equation}%
Indeed, the r.h.s represents the entropy flux across the event
horizon while $\frac{dS_{1}}{d\tau }$ and $\frac{dS_{2}}{d\tau }$
are the contributions to the entropy production rate which depend,
respectively, on the velocity of the infalling matter\ and of the
event horizon:
\begin{eqnarray}
\frac{dS_{1}}{d\tau } &=&P{\int\limits_{\Gamma ^{-}}}d^{3}{\mathbf{r}}d^{3}{%
\mathbf{p}}\frac{ds(\tau )}{d\tau }\frac{dr}{ds}\delta \left(
r-r_{\epsilon
}\right) \hat{\rho}\ln \hat{\rho},  \label{c-1} \\
\frac{dS_{2}}{d\tau } &\equiv &P{\int\limits_{\Gamma ^{-}}}d^{3}{\mathbf{r}}%
d^{3}{\mathbf{p}}\frac{ds(\tau )}{d\tau }\frac{dr_{\epsilon
}}{ds}\delta \left( r-r_{\epsilon }\right) \hat{\rho}\ln
\hat{\rho}.  \label{c-2}
\end{eqnarray}%
As a consequence, $\frac{dS_{1}}{d\tau }$ and $\frac{dS_{2}}{d\tau
}$ can be interpreted as the contributions to the BH entropy
production rate carried respectively by the infalling matter and
the event horizon. Here, $\Gamma ^{-}$ is the sub-domain of phase
space corresponding to the particle falling into the BH. Hence, it
follows that in the above integrals,$\frac{dr}{ds}$ results by
definition negative. Instead, $\frac{dr_{\epsilon }}{ds}$ has not
a definite sign and it can be negative in the case of contracting
event horizons. \ We can also write the above expression in terms
of the kinetic
probability density evaluated at the hypersurface $\sqrt{u_{\mu }u^{\mu }-1}$%
, defined as $\hat{f}(\mathbf{x})\equiv \hat{\rho}/N$. It follows $\hat{\rho}%
\ln \hat{\rho}\equiv N\hat{f}\ln N\hat{f}$. At this point we adopt
a customary procedure in statistical mechanics \cite{Yvon 1969}
invoking the inequality
\begin{equation}
N\hat{f}\ln N\hat{f}\geq N\hat{f}-1
\end{equation}%
and notice that in the sub-domain be in which $F_{{r}{r_{\epsilon
}}}\geq 0$ there results by definition $\hat{\rho}=0$. Hence it
follows that in $\Gamma ^{-}$
\begin{equation}
F_{{r}{r_{\epsilon }}}<0
\end{equation}%
where by construction $\frac{ds(\tau )}{d\tau }>0$. This result
holds independent of the value of $\frac{dr_{\epsilon }}{ds}$.
Next, due to Assumption $\alpha $ and $\beta $ on the surface
$\gamma $ the kinetic distribution function results necessarily
non-zero only in a bounded subset of phase space. Therefore, if $\
\delta $ is an arbitrary infinitesimal,
there exists a bounded subset $\Omega \subset \Gamma ^{-}$ such that $N\hat{f%
}$ results infinitesimal (of order $\delta $) in the complementary set $%
\Gamma ^{-}\setminus \Omega $ (which includes the set of improper points of $%
\Gamma ^{-}$) and moreover
\begin{equation}
P{\int\limits_{\Gamma ^{-}\setminus \Omega }}d^{3}{\mathbf{r}}d^{3}{\mathbf{p%
}}\left\vert F_{{r}{r_{\epsilon }}}\right\vert \delta \left(
r-r_{\epsilon }\right) \left[ N\hat{f}\ln N\hat{f}\right] \sim
O\left( \delta \ln \delta \right) .
\end{equation}%
At least in a subset of $\Omega $, $N\hat{f}$ is by assumption
positive and such that $N\hat{f}>\delta $. Therefore there results
\begin{equation}
0<P{\int\limits_{\Omega }}d^{3}{\mathbf{r}}d^{3}{\mathbf{p}}\left\vert F_{{r}%
{r_{\epsilon }}}\right\vert \delta \left( r-r_{\epsilon }\right)
\equiv M
\end{equation}%
where $M$ is a suitable finite constant. Thus one obtains
\begin{equation}
\frac{dS(\rho )}{d\tau }\geq P{\int\limits_{\Omega }}d^{3}{\mathbf{r}}d^{3}{%
\mathbf{p}}\left\vert F_{{r}{r_{\epsilon }}}\right\vert \delta
\left( r-r_{\epsilon }\right) \left[ Nf-1\right] +O\left( \delta
\ln \delta \right) \label{vel}
\end{equation}%
The first term of the r.h.s of (\ref{vel}) can be interpreted in
terms of the effective radial velocity of incoming particles
\begin{equation}
V_{r}^{eff}\equiv \frac{1}{n_{0}}{\int\limits_{\Omega }}d^{3}{\mathbf{p}}%
\left\vert F_{{r}{r_{\epsilon }}}\right\vert \delta \left(
r-r_{\epsilon }\right) \hat{f},
\end{equation}%
while $n_{0}$ is the surface number density of the incoming
particle distribution function. Finally we invoke the majorization
\begin{eqnarray}
\frac{dS(\rho )}{d\tau } &\geq &\frac{dS}{d\tau }\geq N\inf
\left\{
\int_{\Omega }d^{3}\mathbf{r}n_{0}V_{r}^{eff}\right\}   \nonumber \\
&&-M+O\left( \delta \ln \delta \right)   \label{essep}
\end{eqnarray}%
and notice that, due to the arbitrariness of $\delta ,$ $N$ can
always be chosen sufficiently sufficiently large to satisfy the
inequality $\dot{S}>0.$ We stress that $\inf \left\{ \int
d^{3}\mathbf{r}n_{0}V_{r}^{eff}\right\} $ can be assumed strictly
positive for non isolated BH's surrounded by matter while $\delta
$ is arbitrary. This yields the relativistic H-theorem:

\textbf{THEOREM - }\textit{H-Theorem for the BH classical entropy}

\textit{In validity of the kinetic equation (\ref{Liouville
equation}), boundary conditions (\ref{bc-1}),(\ref{bc-2}) and
regularity conditions defined by Assumptions }$\alpha $\textit{\
and }$\beta ,$\textit{\ the BH classical entropy }$S(\rho )$
\textit{results uniquely defined, according to Eq.(\ref{kinetic
entropy}). Moreover, if the number }$N\gg 1$\textit{\ of classical
point particles of the system }$S_{N}$\textit{\ is assumed
suitably large, there results:}

\begin{equation}
\frac{dS(\rho )}{d\tau }\equiv \frac{dS_{1}}{d\tau }+\frac{dS_{2}}{d\tau }%
_{2}\geq 0,  \label{H-theorem}
\end{equation}%
\textit{where }$\frac{dS_{1}}{d\tau }$ \textit{and}
$\frac{dS_{2}}{d\tau }$
\textit{are respectively the contributions to entropy production rate (\ref%
{c-1}),(\ref{c-2}) due to respectively to infalling matter and the
event horizon. In particular there results }$\frac{dS(\rho
)}{d\tau }=0,$\textit{\ if and only if }$\ n_{0}\equiv 0$\textit{,
being }$n_{0}$\textit{\ the number density on the boundary
surface} $\gamma .$ \textit{Moreover, the
inequality (\ref{H-theorem}) holds even if:}%
\begin{equation}
\frac{dr_{\epsilon }}{ds}<0
\end{equation}%
\textit{(contracting event horizon).}

\section{Conclusions}

Let us briefly analyze the basic implications of this result.
First we notice that the H-theorem here obtained appears of
general validity even if
achieved in the classical framework and under the customary requirement $%
N\gg 1$ (large classical system). Indeed the result applies to BH
having, in principle, arbitrary shape of the event horizon. The
description adopted is
purely classical both for the falling particles (charged or neutral \cite%
{Tessarotto1998,Tessarotto1999,Tessarotto2004,Nicolini2005}) and
for the gravitational field and is based on the relativistic
collisionless Boltzmann equation and/or the Vlasov equation
respectively for neutral and charged particles. A key aspect of
our formalism is the definition of suitable boundary conditions
for the kinetic distribution function in order to take into
account the presence of the event horizon. The expressions for the
entropy and entropy production rate, respectively Eqs.(\ref{Kentropy}) and (%
\ref{entro}), can be used for specific applications to
astrophysical problems. Finally interesting features of the
derivation are that the entropy production rate results
proportional to the area of the event horizon and that the
formalism is independent of the detailed model adopted for the BH.
In particular also the possible presence of an imploding star
(contracting event horizon) is permitted.

\section*{Acknowledgments}
Research developed in the framework of MIUR (Ministero
Universit\'a e Ricerca Scientifica, Italy) PRIN Project \textit{\
Fundamentals of kinetic theory and applications to fluid dynamics,
magnetofluiddynamics and quantum mechanics}, partially supported
(P.N.) by Area Science Park (Area di Ricerca Scientifica e
Tecnologica, Trieste, Italy) and CMFD Consortium (Consorzio di
Magnetofluidodinamica, Trieste, Italy).



\end{document}